\documentclass[twocolumn,preprintnumbers,amsmath,amssymb]{revtex4}
\usepackage{graphicx}
\usepackage{dcolumn}
\usepackage{bm}
\usepackage{color}

\begin{document}

\title{Observation of back-action cancellation\\
in interferometric and weak force measurements}

\author{T. Caniard}
\author{P. Verlot}
\author{T. Briant}
\author{P.-F. Cohadon}
\author{A. Heidmann}

\affiliation{Laboratoire Kastler Brossel, ENS, UPMC, CNRS; case
74, 4 place Jussieu, 75005 Paris, France}
%\homepage{www.spectro.jussieu.fr/Mesure}

%\date{\today}

\begin{abstract}
We experimentally demonstrate a cancellation of back-action noise in
optical measurements. Back-action cancellation was first proposed
within the framework of gravitational-wave detection by dual
resonators as a way to drastically improve their sensitivity. We
have developed an experiment based on a high-finesse Fabry-Perot
cavity to study radiation-pressure effects in ultra-sensitive
displacement measurements. Using an intensity-modulated intracavity
field to mimic the quantum radiation-pressure noise, we report the
first observation of back-action cancellation due to a destructive
interference between radiation-pressure effects on both mirrors of
the cavity. We have observed a sensitivity improvement by a factor
larger than 20 both in displacement and weak force measurements.
\end{abstract}

\pacs{04.80.Nn, 42.50.Lc, 03.65.Ta}

\maketitle

{\it Introduction.}-- The unique sensitivity of optical
interferometry is used in many high-sensitivity measurements,
including gravitational-wave interferometers
\cite{Bradaschia90,Abramovici92}, optical transduction of the
gravitational signal in Weber bars \cite{Conti03}, monitoring of a
micro-mechanical resonator \cite{PRL06,Sylvain}, or for weak force
detection such as AFM \cite{Rugar}. Quantum noise of light however
induces fundamental limits: current gravitational-wave
interferometers are already confronted to the shot-noise limit at
high frequency and the next generation of interferometers
\cite{Ligo2} and dual detectors with optical readout
\cite{Cerdonio2001} will be confronted to the back-action noise
through the radiation pressure on the test masses. A number of
sophisticated quantum noise reduction schemes have been proposed to
overcome the related {\it Standard Quantum Limit} (SQL): use of
squeezed light sent into the interferometer \cite{caves,Reynaud},
use of quantum correlations induced by radiation pressure between
phase and intensity fluctuations \cite{kimble}, quantum locking of
mirrors \cite{courty} or detuning of the signal recycling cavity
\cite{buonanno}. Recent results have also shown that the specific
geometrical design of dual resonators \cite{Cerdonio2001} allows for
their sensitivity to be strongly improved by a back-action noise
cancellation effect, due to a destructive interference between
radiation-pressure effects on both resonators
\cite{dual-resonators}.

We present in this paper the first experimental demonstration of
such a back-action cancellation. We have developed an experiment to
study radiation-pressure effects in a high-finesse Fabry-Perot
cavity. A dual optical injection system allows us to mimic the
quantum radiation-pressure noise by using a noisy classical
intensity modulation \cite{PRLAustraliens}. We have observed a
classical back-action cancellation and we demonstrate a drastic
sensitivity improvement beyond the limit imposed by usual
radiation-pressure noise, for the measurements of both a cavity
length variation and of a weak force applied to one cavity mirror
used as a mechanical transducer.

{\it Back-action cancellation in a dual resonator sensor.}-- We
consider a probe laser beam sent into a single-port lossless optical
cavity. We study the response of the system to a variation $\delta
L_{\rm sig}$ of the cavity length. The displacement fluctuations of
the front and end mirrors are denoted $\delta X_{\rm f}$ and $\delta
X_{\rm e}$, respectively. The cavity length fluctuations $\delta L$
are then given by
\begin{equation}
\delta L=\delta X_{\rm e}-\delta X_{\rm f}+\delta L_{\rm sig}.
\label{eq:L}
\end{equation}

We now determine the quantum-limited sensitivity of the
optomechanical sensor in the measurement of $\delta L_{\rm sig}$.
When the cavity is at resonance, the fluctuations $\delta q^{\rm
out}\left[\Omega\right]$ of the phase quadrature of the reflected
beam at a frequency $\Omega$ are given by \cite{courty}
\begin{equation}
\delta q^{\rm out}[\Omega]=\delta q^{\rm
in}[\Omega]+\frac{16\mathcal F}{\lambda} \sqrt{I^{\rm in}} \delta
L[\Omega],\label{eq:deltal}
\end{equation}
where $\lambda$ is the laser wavelength, ${\mathcal F}$ the cavity
finesse, $I^{\rm in}$ the incident power (normalized as a photon
flux), $\delta q^{\rm in}$ the phase-quadrature fluctuations of the
incident beam, and we have assumed $\Omega$ much smaller than the
cavity bandwidth.

Working at low temperature, the ultimate displacement noise is the
back-action noise due to the radiation pressure exerted by the
intracavity field on both mirrors: $F_{\rm rad}(t)=2\hbar k I(t)$,
where $k=2\pi/\lambda$ is the field wavevector and $I(t)$ the
intracavity intensity. The response of the mirrors to radiation
pressure is then described by linear response theory:
\begin{equation}
\delta X_{\rm e}[\Omega]-\delta X_{\rm f}[\Omega] =\left( \chi_{\rm
e}[\Omega]+\chi_{\rm f}[\Omega]\right)F_{\rm rad}[\Omega],
\label{eq:ChiFrad}
\end{equation}
where $\chi_{\rm e}$ and $\chi_{\rm f}$ are the mirrors'
mechanical susceptibilities. The sign change in the right-hand
part of Eq. (\ref{eq:ChiFrad}) is due to the opposite direction of
the radiation-pressure force on both mirrors.

\begin{figure*}
\includegraphics[width=13 cm]{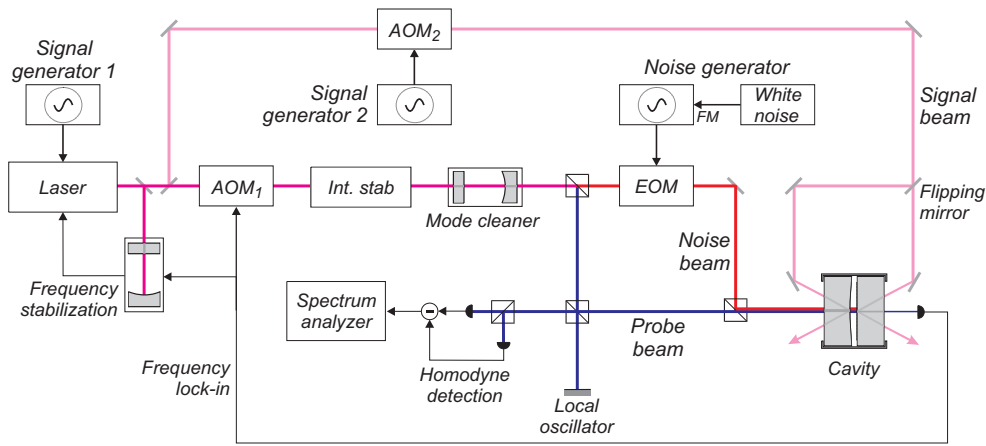}
\caption{\label{fig:setup}Experimental setup. The high-finesse
cavity is composed of two low-loss cylindrical mirrors. The
frequency- and intensity-stabilized laser source is locked on the
cavity resonance via an acousto-optic modulator (AOM$_1$). Two
beams are sent into the cavity: a noise beam which is
intensity-modulated by an electro-optic modulator (EOM) in order
to apply a radiation-pressure noise on the mirrors, and a weaker
probe beam whose reflected phase is measured by homodyne detection
to monitor the mirror displacements. The signal in the measurement
can either be an optical length variation of the cavity mimicked
by a frequency modulation of the laser (signal generator 1), or a
weak force applied through the radiation pressure of an
intensity-modulated signal beam onto the front or end mirror of
the cavity (signal generator 2). For simplicity, polarizing
selective elements are not shown.}
\end{figure*}

For a coherent state of light, the incident phase and intensity
quadratures are uncorrelated and their noise spectra are normalized
to 1. The sensitivity for a measurement of $\delta L_{\rm sig}$
therefore stems from a trade-off between phase and intensity noises.
At low power, phase noise prevails and the sensitivity is given by
$\delta L_{\rm shot}[\Omega] =\lambda/\left(16{\mathcal
F}\sqrt{I^{\rm in}}\right)$, whereas radiation-pressure effects
prevail at higher power with a corresponding sensitivity
\begin{equation}
\delta L_{\rm ba}[\Omega]=\left|\chi_{\rm e}[\Omega]+\chi_{\rm
f}[\Omega]\right|\frac{4\hbar k{\mathcal F}}{\pi} \sqrt{ I^{\rm
in}}\label{eq:deltaxba}.
\end{equation}
In the dual sphere antenna recently proposed for wideband
gravitational-wave detection \cite{Cerdonio2001}, two spherical
masses are nested together with only a small gap in-between whose
length can be monitored by an optical cavity. The fundamental
modes of both spheres have different resonant frequencies and
their response to a gravitational wave are out-of-phase for
intermediate frequencies: the global response is then enhanced
when measuring the gap in-between. In contrast, as
radiation-pressure forces have opposite directions, one gets a
destructive interference between back-action effects on both
resonators: the global back-action noise $\delta L_{\rm ba}$ is
almost cancelled at the anti-resonance frequency $\Omega$ for
which the sum $\chi_{\rm e}+\chi_{\rm f}$ vanishes. The
measurement is then mainly limited by the shot noise $\delta
L_{\rm shot}$ and the optimization of the incident light power
leads to the sensitivity
\begin{equation}
\delta L_{\rm min}[\Omega]=\sqrt{\hbar \left| \chi_{\rm
e}[\Omega]+\chi_{\rm f}[\Omega] \right|}\label{eq:deltaxmin},
\end{equation}
similar to a SQL sensitivity taking into account both resonators
and their destructive interference. One then gets a drastic
sensitivity improvement in the intermediate frequency band
\cite{dual-resonators}: at the anti-resonance frequency, the real
parts of the susceptibilities exactly compensate for each other,
leaving only the imaginary part in expression
(\ref{eq:deltaxmin}). The sensitivity then reaches the ultimate
quantum limit of the measurement \cite{Reynaud}.

{\it Experimental setup.}-- Our experimental setup is based on a
very-high finesse cavity made of two 1-inch high-quality
cylindrical mirrors with very low losses: we have obtained a
cavity finesse $\mathcal F=230\,000$. The cavity is very short
($L=0.25\,{\rm mm}$) in order to have a large cavity bandwidth
($\Omega_{\rm c}/2\pi=1.3\,{\rm MHz}$) and to reduce the influence
of laser frequency noise. It is operated in vacuum to increase the
mechanical quality factors.

The light beam entering the cavity is provided by a Ti:Sa laser
working at 810 nm, frequency-stabilized on an external reference
cavity. The light beam is also intensity-stabilized and spatially
filtered by a mode cleaner. The phase fluctuations of the reflected
beam are monitored by a homodyne detection. For an incident power of
50 $\mu$W, one gets a quantum-limited sensitivity of $2.7\times
10^{-20}\,{\rm m}/\sqrt{{\rm Hz}}$ at frequencies above 200 kHz.

In order to mimic the quantum fluctuations of radiation pressure, we
have developed a dual-beam injection system: a second light beam
(noise beam in Fig. \ref{fig:setup}, with a $300\,\mu {\rm W}$
power, cross-polarized with the probe beam) is intensity-modulated
with an electro-optic modulator (EOM) before entering the cavity to
produce a classical intracavity radiation-pressure modulation. The
EOM is driven by a high-frequency generator synchronized with the
spectrum analyzer used in the homodyne detection, and
frequency-modulated by a $600$-Hz wide gaussian white noise. As
compared to the $10$-Hz resolution bandwidth of the spectrum
analyzer, the resulting incident intensity fluctuations thus appear
as a white noise which properly mimics -though at a higher level-
the quantum fluctuations of the incident intensity.

As the experiment requires a perfect isolation of the phase of the
probe beam with respect to the intensity of the noise beam, we have
carefully eliminated unwanted optical reflections and residual
birefringence of the high-finesse cavity. The optical rejection of
the double-beam system is higher than 35 dB: in our experimental
conditions, observable effects of the noise beam are therefore
necessarily mediated by intracavity radiation pressure.

An additional intensity-modulated beam (signal beam in Fig.
\ref{fig:setup}, with a $1\,{\rm W}$ power) can be used to apply a
weak force ($\simeq$ 1 nN) upon one of the mirrors through
radiation pressure: a flipping mirror allows to actuate either the
front or end mirror of the cavity. The beam spot can also be swept
over the mirror surface in order to map the vibration profile of
each mode \cite{Briant03}.

\begin{figure}
\includegraphics[width=7 cm]{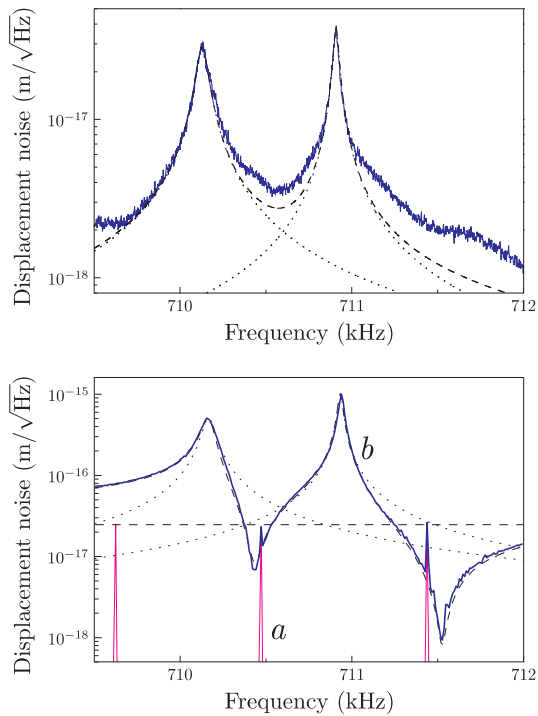}
\caption{\label{fig:thermal}Top: thermal noise spectrum of the
cavity in the vicinity of a particular doublet. Each peak is
related to a specific vibration mode of one mirror. The dashed
line is a double lorentzian fit corresponding to the quadratic sum
of individual mirrors' thermal noises (dotted lines).
Bottom:~measurement of an optical length variation of the cavity
produced by a modulation of the laser frequency. Curve {\it a}
shows the monochromatic signal at different modulation
frequencies, with the noise beam off. In presence of
radiation-pressure noise (curve {\it b}), the signal is still
observable in the dips associated with back-action cancellation.
Dashed curves are theoretical fits of the signal and
radiation-pressure noise, and dotted curves are the expected
individual radiation-pressure noise spectra.}
\end{figure}

{\it Observation of radiation-pressure cancellation.}-- Cancellation
effects are expected in our system since both mirrors have similar
sizes and present acoustic modes at nearly equal resonance
frequencies. The overall mechanical response thus appears as a set
of resonant doublets. Near one particular doublet, the motion of
each mirror is mainly ruled by its resonant mode, with a lorentzian
mechanical susceptibility given by
\begin{equation}
\chi_i[\Omega]=\frac1{M_i\left(\Omega_i^2-\Omega^2-i\Omega\Omega_i/Q_i\right)}+\chi_i^{(0)},
\label{eq:chi_ni}
\end{equation}
where the index $i=\{\mathrm{e,f}\}$ stands for the mirror,
$\Omega_i$ is the resonance frequency of the mode, $M_i$ its
effective mass and $Q_i$ its mechanical quality factor. The
susceptibility $\chi_i^{(0)}$ describes the out-of-resonance
response of all other mechanical modes of the mirror, assumed
constant near the doublet. Due to the geometry discrepancy, the
two resonance frequencies $\Omega_i$ are slightly shifted. As the
mechanical responses are out-of-phase for intermediate
frequencies, one expects back-action cancellation in that
frequency domain.

Figure \ref{fig:thermal} shows the thermal noise spectrum observed
in the vicinity of a particular doublet. Since thermal noises of
both mirrors are not correlated, their contributions are simply
added as shown from the double lorentzian fit. Excess noise is due
to a neighboring doublet. We have checked by a selective optical
actuation of each mirror that the lower-frequency resonance is due
to the front mirror whereas the higher-frequency one is due to the
end mirror. The fit yields the following parameters: $\Omega_{\rm
f}/2\pi=710.1\,\mathrm{kHz}$, $\Omega_{\rm
e}/2\pi=710.9\,\mathrm{kHz}$, $M_{\mathrm{f}}=0.64\,\mathrm{g}$,
$M_{\mathrm{e}}=0.84\,\mathrm{g}$, $Q_{\mathrm{f}}=10\,500$, and
$Q_{\mathrm{e}}=21\,500$. The discrepancy between the two
effective masses $M_{\mathrm{f}}$ and $M_{\mathrm{e}}$ can be
attributed to the different spatial overlap between the cavity
field and the vibration modes \cite{Briant03}.

We now turn the noise beam on and set its intensity-noise level in
order for the mirror displacements induced by radiation pressure to
be well above the thermal noise at every frequency. Curve {\it b} of
Fig. \ref{fig:thermal} (bottom) shows the resulting mirror
displacement spectrum, obtained by scanning the modulation frequency
of the noise beam over the whole doublet, with a 10-Hz resolution
bandwidth of the spectrum analyzer: one gets a clear cancellation of
radiation-pressure effects between the two resonances as compared to
the individual responses of each mirror (dotted lines deduced from
the characteristics of the thermal noise spectra). The coherence
between both radiation-pressure white noises driving the mirrors and
their out-of-phase responses make them enter a 'common-mode' motion
no longer observable by the probe beam. Back-action noise due to
radiation pressure is then transferred to the common-mode motion.
Another cancellation dip is observed at a higher frequency, where
the out-of-phase response of the two resonant modes compensates the
in-phase background response of all other modes.

To demonstrate the sensitivity improvement in length measurements,
we generate a monochromatic modulation $\delta \nu_{\rm sig}$ of
the laser optical frequency $\nu$, equivalent to an apparent
variation $\delta L_{\rm sig}$ of the cavity length with $\delta
L_{\rm sig}/L= \delta \nu_{\rm sig}/\nu$. Curve {\it a} of Fig.
\ref{fig:thermal} (bottom) shows the effect of a small modulation,
corresponding to optical length variations in the $10^{-17}\,{\rm
m}$ range, successively applied at different frequencies, with the
noise beam off. In presence of radiation-pressure noise (curve
{\it b}), the signal is no longer observed except at frequencies
where back-action cancellation occurs. As compared to the
individual radiation-pressure contributions (dotted lines), one
gets a sensitivity improvement by a factor 25.

{\it Weak-force sensitivity enhancement.}-- Back-action
cancellation may also be useful to other optical measurements such
as weak force detection \cite{Rugar}. For an optomechanical
resonator used as a weak force sensor with a mechanical
susceptibility $\chi[\Omega]$, the SQL of the displacement
measurement leads to a related quantum limit for the force
\cite{SQLforce}
\begin{equation}
\delta F_{\rm
SQL}[\Omega]=\sqrt{\hbar/\left|\chi[\Omega]\right|}\label{eq:Fsql}.
\end{equation}

Consider now the sensor as the end mirror of a Fabry-Perot cavity
(with $\chi=\chi_{\mathrm{e}}$), with an almost identical one as
front mirror. Since the latter is unsensitive to the weak force, one
can take advantage of the back-action cancellation inside the
optical cavity to reach a sensitivity
\begin{equation}
\delta F_{\rm min}[\Omega]=\frac{\sqrt{\hbar \left| \chi_{\rm
e}[\Omega]+\chi_{\rm f}[\Omega] \right|}}{|\chi_{\rm
e}[\Omega]|},\label{eq:Sf}
\end{equation}
no longer limited by $\delta F_{\rm SQL}$ at anti-resonance
frequencies.

\begin{figure}
\includegraphics[width=7 cm]{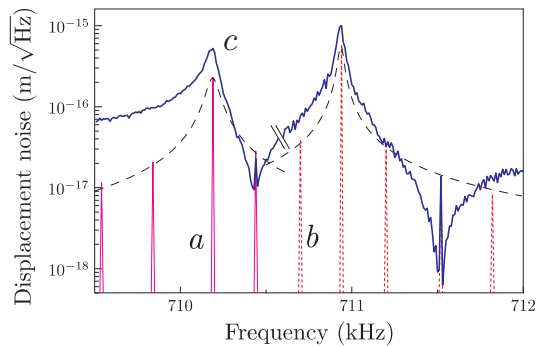}
\caption{\label{reponse_force.eps} Measurement of a weak force
applied at different frequencies by the signal beam, either to the
front (peaks {\it a} around the first resonance) or end (peaks {\it
b} around the second resonance) mirror. Curve {\it c} is obtained in
presence of radiation-pressure noise, with the force applied to the
front mirror for frequency less than 710.6 kHz, and to the end
mirror at higher frequency: the force is observable in the dips
associated with back-action cancellation. Dashed curves are fits of
both mirrors' dynamical responses to the force.}
\end{figure}

As an illustration, we use our cavity to measure a weak force
produced by the intensity-modulated signal beam. This signal is
set 5 dB below the back-action noise of the intracavity noise beam
at the mechanical resonance frequency of the sensor. The frequency
of the monochromatic force applied with the signal beam is scanned
from 709.5 to 712 kHz: Fig. \ref{reponse_force.eps} shows the
experimental spectra obtained with (curve {\it c}) and without
(curves {\it a} and {\it b}) the noise beam, when the weak force
is either applied onto the front (four first peaks) or end (five
last peaks) mirror of the cavity. At the vicinity of back-action
cancellation frequencies, curve {\it c} clearly shows that the
weak force, otherwise unobservable, can be measured with a
signal-to-noise ratio up to the order of 10.

{\it Conclusion.}--  We have demonstrated classical back-action
cancellation, which can be considered as a proof-of-principle
demonstration of quantum back-action cancellation
\cite{dual-resonators} as quantum-limited interferometry can be
understood in purely classical terms \cite{PRLAustraliens}. Though
classical intracavity radiation-pressure effects have already been
demonstrated with specially designed mirrors
\cite{Australiens,Nature} or suspended mirrors \cite{Mavalvala},
this is the first demonstration of such effects with the much
stiffer resonators provided by the internal vibration modes of
standard fused silica mirrors. The correlations between the phase of
the probe beam and the intensity of the noise beam could be extended
to the quantum level by further experimental progress and cryogenic
operation of the cavity, allowing for radiation-pressure induced QND
measurement of light intensity \cite{QND} and other quantum optics
experiments.

We gratefully acknowledge Jean-Marie Mackowski and his group at
the Laboratoire des Mat\'eriaux Avanc\'es for the optical coating
of the low-loss mirrors. This work was partially funded by EGO
(collaboration convention EGO-DIR-150/2003 for a study of quantum
noises in gravitational wave interferometers) and by the
Integrated Large Infrastructures for Astroparticle Science (ILIAS)
of the Sixth Framework Program of the European Community.

\end{document}